# Propagation of short lightpulses in microring resonators: ballistic transport versus interference in the frequency domain

A. Driessen[1], D.H. Geuzebroek[1], E.J. Klein[1], R. Dekker[1], R. Stoffer[2] and C. Bornholdt[3]

[1]*Integrated Optical Micro Systems, MESA$^+$ Institute, University of Twente, P.O. Box 217, 7500 AE Enschede, The Netherlands*
[2]*Phoenix, P.O. Box 545, 7500 AM Enschede, The Netherlands*
[3]*Fraunhofer Institute, Heinrich Hertz-Institute, Einsteinufer 37, 10587 Berlin, Germany*

**Abstract**
The propagation of short lightpulses in waveguiding structures with optical feedback, in our case optical microresonators, has been studied theoretically and experimentally. It appears that, dependent on the measurement set-up, ballistic transport or interference in the time domain of fs and ps laser pulses can be observed. The experiments are analyzed in terms of characteristic time scales of the source, the waveguide device and the detector arrangement and are related to Heisenberg's uncertainty principle. Based on this analysis a criterion is given for the upper bitrate for error free data transmission through optical microresonators.

## 1. Introduction

Optical microring resonators (MR) [1], [2] are attractive photonic structures that can be applied in optical communication and optical sensing as filter, space-switch, modulator, sensor and all-optical data processing device. They are small with a diameter in the order of tens of μm and can be clustered in arrays leading eventually to Very Large Scale Integrated (VLSI) photonics [3]. With increasing bitrates in optical communication, currently up to 40 Gbit/s, a thorough understanding of the response to short pulses is desired. Feeding a MR with constant input intensity, the resulting Fabry-Perot like fringes at the output ports are well understood. If, however, the pulse duration is reduced well below the roundtrip time $\tau_r$, two basic possibilities should be considered: either ballistic transport [4] or interference in the frequency domain (IFD)[5]. In the latter case, the output spectrum is equal to the input spectrum times a transmission function [6]. In the case of ballistic transport, however, the multiple paths in the resonator lead to a train of reduced replica of the incoming pulse at the output ports.

In the following we will analyze numerically and experimentally the response of MRs to short pulses. These pulses are assumed to be transform limited. In this case we will show that the qualitative features of the resulting spectral response are largely dependent of three characteristic time scales: the duration $t_p$ of the incoming pulse, the roundtrip time $\tau_r$ and the inherent time uncertainty $\Delta t$ in the measurement set-up consisting of spectrometer and detector. The paper is organized as follows. After the brief introduction some theoretical considerations are given with respect to the occurrence of IFD and the properties of a MR. Thereafter, it is shown on the basis of the Heisenberg uncertainty principle (HUP) that the resolution of a wavelength selective device is related to a principle residence time uncertainty of the photon in that device. In the third section a number of experiments and numerical calculations are given that show the spectral response of MRs for certain choices of the three characteristic timescales. The paper is ended with discussion and conclusions including an estimate of the bit rate limits of MRs used as filters in multi-wavelength optical communication.

## 2. Theoretical considerations
*2.1 Interference in the frequency domain*

Light can be considered as electromagnetic radiation as described by the equations of Maxwell or as a quantum object, the photon [7]. The two approaches lead in most cases to identical results. Events, however, where single photons are generated or absorbed can not be described by Maxwell's equations. In this paper both approaches are used in order to relate important parameters to each others. Interference of light is a natural consequence of the wave-like approach in the Maxwell theory; in quantum mechanics it is a manifestation of coherent superposition of probability amplitudes for undistinguishable paths [8]. Interference in the space domain, where enhancement or extinction occurs at certain positions in space, is since Young's double slit experiment a well-known phenomenon. Interference in the frequency domain is an analogue effect where the enhancement and extinction occurs at certain frequencies within the spectrum of the light source. The three interference phenomena, in space, time or frequency come into play if one considers vectorial fields that eventually have to be converted to light intensities in the Maxwell picture or, in QM, to probabilities to encounter a photon within a given frequency range at a given place and time.

IFD can be observed when the **E**-field generated by a photon source has a rapidly in time varying envelop and propagates to the detector -position $x$ - by different trajectories with a trajectory and wavelength dependent delay and phase. For certain frequencies the time varying field amplitudes arriving by different paths at the detector have the same phase constant and contribute coherently to the light intensity; for others there is at least partly destructive interference. As one is interested in the probability to detect a photon within a certain frequency range $\Delta \nu$, one has to convert the time dependence of the **E**-field into a frequency dependence with the aid of a Fourier transformation:

$$\mathbf{E}(\mathbf{x},\nu,t_0,\Delta t) = C \int_{t_0-\Delta t/2}^{t_0+\Delta t/2} \mathbf{E}(\mathbf{x},t) e^{i\nu t} dt \qquad (1)$$

where C is a constant. Two important observations can be made. First, besides the time $t_0$ of the detection event also the boundary of the integral, i.e. the duration $\Delta t$, is of great importance. That duration, which in general is not the coherence time of the photons of the source or the pulse duration in the case of pulsed sources, is related to a characteristic time of the measurement arrangement that may include besides a detector a wavelength selecting device like a spectrometer. Secondly it appears that fields **E**(**x**,t) at the same position $x$ but at different times can interfere as long as it happens within $\Delta t$. The probability of finding a photon within a certain frequency range is proportional to the square of $|\mathbf{E}(\mathbf{x},\nu,t_0,\Delta t)|$.

*2.2 The optical microring resonator*

The MRs in this study consist of an integrated optics waveguide ring coupled to two adjacent linear waveguides serving as input and output port, see Fig. 1(a). Light enters at $I_{in}$ and couples partly at the left coupler to the ring, the other part continues to the output port $I_{through}$. The light in the ring propagates to the right coupler, where a part is dropped to port $I_{drop}$. The other part follows the ring to the left coupler where part of the field is coupled to $I_{through}$ and the remaining keeps propagating in the ring. In both cases interference occurs with the new incoming light. If the roundtrip phase is a multiple of π, the resonator is on resonance and constructive interference occurs inside the ring and destructive at $I_{through}$. After the first round, light propagation inside the ring continues with decreasing amplitude. The resulting spectral response is similar to that of a Fabry-Perot interferometer and consists of periodic dips and peaks at the through- and drop port respectively, see Fig. 1(b). These dips and peaks occur at resonance. In that case the phase of the light dropped from the resonator to the through port has a phase shift of π with respect to the field running directly from the input to the through port. At a wavelength exactly half between the resonances this phase shift is zero.

The functional behavior of a MR is characterized by its Free Spectral Range (FSR), i.e. the wavelength separation between neighboring resonances; the width at half maximum of the resonance peak, $\Delta\lambda_{-3dB}$, and the finesse $F = FSR/\Delta\lambda_{-3dB}$. The average number of roundtrips $m$ of photons is related to the finesse by: $m = F/(2\pi)$; the average residence time in the resonator is $m\tau_r$.

*2.3 The role of the Heisenberg uncertainty principle in wavelength selecting structures*

The question remains, how the range of the integral $\Delta t$ in Eq. (1) is related to experimental parameters. A good starting point for further analysis is the inherent connection between the ranges in frequency and time, $\delta v$ and $\delta t$ respectively. Both, Fourier analysis and the Heisenberg Uncertainty Principle (HUP) state that there is a minimum value of the product of the uncertainties in frequency and time:

$$\delta v \delta t \geq \frac{1}{4\pi} \quad (2)$$

For a photon with energy $E = hv$, (h is the Planck constant) Eq. (2) means that the product of the uncertainties in energy and a critical time or duration has a minimum [9]. For a transform limited pulse Eq. (2) can be written as:

$$\delta v \delta t = k/4\pi \quad (3)$$

with the value of $k$ depending on he definition of the uncertainties and the specific pulse shape, for example $k = 5.54$ for a Gaussian pulse [10]. The frequency $v$ is related to the wavelength $\lambda_0$ and the speed of light $c$, both in vacuum by:

$$v = c/\lambda_0 \quad (4)$$

In addition, the uncertainty in time can be related to an uncertainty in length:

$$\delta l = c\delta t \quad (5)$$

where $\delta l$ is a fundamental uncertainty in the optical path length of the photon. With Eqs. (2) written as equality ($k=1$) and (4) we obtain for the fundamental limit in the uncertainty in wavelength $\delta\lambda_0$:

$$\delta\lambda_0 = \lambda_0^2/(4\pi c\delta t) \quad (6)$$

This relation is obtained by applying the HUP to a photon and states that by maximizing the uncertainty in time one obtains the smallest uncertainty in wavelength. The fundamental physical principle of wavelength selecting structures therefore is the introduction of a fundamental time uncertainty in the lightpath or alternatively, with Eq. (5) a fundamental uncertainty in the length of the optical path.

The validity of Eq. (6) can be easily understood in the case of a Fabry-Perot resonator, where the principal time uncertainty arises from the uncertainty in the number of times a single photon is oscillating forth and back before leaving the resonator. Only known is the average number $m$. The time uncertainty in a FP therefore is given by:

$$\delta t = 2mnd/c = Fnd/\pi c \quad (7)$$

with $n$ the index of refraction and $d$ the cavity length. Combining Eq. (2) and (6) one gets for the frequency resolution of a FP resonator:

$$\delta v = \frac{c}{4Fnd} \quad (8)$$

This is within a factor 2 the minimum resolvable bandwidth as derived by Hecht [11]. The factor 2 arises from the ambiguity of the value of $k$ in Eq. (3). Born and Wolf [12] derived Eqs. (2), and (6) directly without using the HUP. A simple geometrical analysis shows that with Eq. (6) the same expression for the resolving power of a grating can be derived as given by Born and Wolf. One may generalize the results obtained for a FP cavity and a grating by stating that Eq. (6) is a valid expression for the fundamental limit in resolution of any wavelength selecting devices.

*2.4 The relevant characteristic time durations*

Considering a general scheme for experiments with short light pulses in optical MRs, see Fig. 2, the relation between the above mentioned characteristic durations can be further established. The lightsource can be described by a single duration $t_p$, the pulse length, if the distance between successive pulses is assumed to be very large. This condition is fulfilled in the experiments described below as the short pulses belong to mode-locked trains where the subsequent pulses are effectively independent since the mode-locked laser is very much larger than the MR. The MR can be characterized by the roundtrip time $\tau_r$. By intuition, one already expected that the range of the integral $\Delta t$ in Eq. (1) should be related to the wavelength resolution of the measuring set-up. If one identifies this duration with the time uncertainty $\delta t$ in the HUP in the form of Eq. (6), this relation is given explicitly. The measurement set-up consists evidently not only of the wavelength selecting device but also of a photon detector with electronics having its own characteristic duration $t_D$ for the detection of a photon. The identification between $\Delta t$ and $\delta t$ is only valid if $t_D \gg \delta t$. In the opposite case, when $t_D \ll \delta t$, the identification has to be made between $\Delta t$ and $t_D$, that means that spectroscopic details are smoothened with increasing detector time resolution.

Being an interference effect, IFD can be expected if there is no *which-way* information. Evidently, this information is only possible if the number of roundtrips of a photon in the MR is uniquely defined, that means that $t_p < \tau_r$ and $\Delta t < \tau_r$. Under this condition, ballistic transport of the lightpulses and generation of a pulse train in the MR is expected. In the following we will describe experimental and numerical results on the response of MRs to ps and fs pulses with different measurement set-ups where ballistic transport as well as IDF could be observed. Also measurements are presented for the intermediate situation.

## 3. Numerical and experimental results

The experiments were performed during recent years on several integrated optical MRs which were designed, realized and characterized by cw spectroscopy at MESA[+]-IOMS. The short pulse measurements were carried out in collaboration with colleagues at MESA[+], HHI Berlin and RWTH Aachen. The devices have been realized in SiON technology [13] with vertical as well as lateral coupling between ring and port waveguides [2]. The FSR was always a few nm and $\tau_r$ in the order of ps. An overview of the MRs with some relevant parameters is given in Table I.

| # MR | material | radius [μ] | FSR [nm] | F | $\tau_r$ [ps] | references with Details of MR |
|---|---|---|---|---|---|---|
| 1 | SiON | 25 | 9 | 20 | 1 | [2],[14] |
| 2 | SiON | 100 | 2 | 10 | 4 | [15] |
| 3 | SiON | 50 | 4.3 | 10 | 1.9 | [16] |
| *Table I: Overview of relevant parameters of the MRs used in the experiments* | | | | | | |

*3.1 Experimental and numerical results with fs pulses in a MR*

In a first experiment propagation of fs lightpulses in an integrated optical MR with a diameter of 50 μm is studied with numerical as well as experimental means. The microresonator (MR1, Table I) is made in silicon-based technology in a lateral coupling geometry. The two adjacent port waveguides are mode matched to the resonator ring, $n_{eff}$ ~

1.67 with a field coupling constant ~ 0.3. The numerical calculations based on this geometry are performed with a two dimensional Time Domain Beam Propagation Method (TDBPM) based on the effective index method [17] and exhibit the same qualitative features as those by Hagness et al.[18]. Fig. 3 shows the field response to a 20 fs pulse at the output ports for the first 22 pulses. Bars indicate the pulses obtained at the drop port. The pulse train obtained at the drop port is very similar to the one of the through port with exception of the missing first large pulse.

For the experiments, Gersen et al. [14] used a photon scanning tunneling microscope (PSTM) to track the propagation of pulses of 123 fs duration at 1300 nm with a spatial resolution of ~ 100 nm. They obtained the field amplitude and optical phase as a function of time with fs resolution, see Fig. 4. Clearly the ballistic propagation of the fs light pulse can be seen in the ring as well as in the port waveguides. When comparing the experimental field amplitude and phase obtained at incremental time with TDBPM calculations, essential agreement is found.

When converting the **E**(t) data to the frequency domain, a Fourier transform has to be carried out by use of Eq. (1). Crucial for this is the selection of the duration $\Delta t$ in the integral or in the experiment the wavelength resolution of the detection set-up. In the case $t_p << \tau_r << \Delta t$, there is no possibility to know *which way* the photons have gone through the system, it is principally unknown whether they traveled once, twice, n-times or even not at all around the ring resonator. Accordingly, one would expect to observe interference effects. Using the electric field data obtained by TDBPM and Eq. (1) one ends up with the cw transfer function of the resonator multiplied by the input pulse spectrum for the through port, Fig. 5(a), as well as for the drop port, Fig. 5(b). It is by IFD that one obtains the expected spectra identical to cw excitation with a broadband source with a frequency distribution equal to the pulse source. The same conclusions are experimentally confirmed by Geuzebroek and al. [15] who compared in an experiment with a similar MR (MR2 in Table I) wavelength spectra measured with a spectrometer with resolution 0.1 nm, corresponding to $\Delta t$ in the order of 10 ps. Fig. 6 shows the spectra obtained with a cw laser source or alternatively with a 200 fs laser.

If $t_p < \Delta t < \tau_r$ or $\Delta t < t_p < \tau_r$ *which-way* information is completely available. The time of detection determines uniquely the number of roundtrips the photon has carried out within the resonator. In consequence no IFD of photons traveling different paths can be expected. As is shown by Gersen et al. [19], a simple Fourier transform of the experimental PSTM data results in the wavelength spectrum of a fs pulse, which in the case of the data in Fig. 4 will evidently show no spectral details related to the resonances. Using alternatively the calculated field E(t) of the first leading pulse at the through port obtained by TDBPM, see Fig 2, one also gets essentially a replica of the source pulse. With other words, the light pulse behaves as a quasiparticle which undergoes ballistic transport. The absence of spectral detail can also be explained by Eq. (2). As the measuring time is short, no fine structure in the spectra can be expected. The necessary condition for observing MR induced resonance features in the through- or drop port is $\tau_r < \Delta t$ in contradiction with the assumption made for this case.

### 3.2 Experimental results with ps pulses in a MR

In the two extreme cases considered above, either ballistic transport ($\tau_r >> t_p$ and $\tau_r >> \Delta t$) or IFD ($\tau_r < \Delta t$) are observed. For applications in wavelength division multiplexing (WDM) networks the wavelength selective properties of MRs are exploited. In this case the condition for IDF should be fulfilled. With increasing bitrate the effective $t_p$ and $\Delta t$ will decrease leading eventually to the transition to ballistic transport without the desired wavelength selectivity. In order to study the transition regime in more detail we performed measurements (for preliminary results see [20]) of the response of a tunable microresonator to a 40 GHz modelocked ringlaser with nearly Fourier transform limited pulses of about 2 ps duration and a time separation of 25 ps between successive pulses. The resonator, MR3 in Table 1 [16], has a radius of 50 μm, a FSR of 4.3 nm and a roundtrip time of 1.9 ps so that $\tau_r \sim t_p$.

The experimental results are presented in Fig. 7 a-f. In the first row of Fig. 7, the incoming pulse is measured with an optical sampling system with a bandwidth of 700 GHz, with an effective $t_D$ of ~1 ps (Fig. 7.a). In addition the wavelength response is obtained with the aid of a spectrometer with a resolution of 0.01 nm corresponding to Δt in the order of 100 ps (Fig. 7.b). The pulse has a nearly Gaussian shape with a width of circa 2.3 ps. In the wavelength domain an envelope is seen with spikes. The envelope is the Fourier transform of the time spectrum of a single pulse resulting in a width in the wavelength spectrum of circa 3.5 nm. The spikes have a fixed distance in wavelength of Δλ= 0.32 nm corresponding to 40 GHz, the repetition rate of the mode-locked laser source.

In the next experiment the same pulses are directed to the input port of MR3 and detected at the drop port. In the first set of measurements the MR was thermally tuned such that the maximum of the wavelength spectrum of the incoming pulse was at off-resonance, see Fig. 7.c and d. The wavelength spectrum has been obtained with the same spectrometer as used for the pulse of the laser source in 7.b. That means that $\Delta t > t_p$ and also $\Delta t > \tau_r$, so that the condition for IDF is fulfilled. In this case the wavelength spectrum is the product of the incoming spectrum, Fig. 7.b, with the transmission function of the MR, which is given in the insert of Fig. 7.d. As a result the envelope of the incoming pulse is broadened (3 dB width 4.5 nm) as the transmission function of the MR reduces the intensity at the central wavelength and amplifies the wings. With a broadened wavelength spectrum the resulting pulse width in the time domain is reduced to 1.9 ps as determined with the high speed photo detector, see Fig. 7.c. Besides the leading pulse with high intensity a series of smaller pulses is visible indicating ballistic transport of the lightpulse.

In the second set of measurements the MR is tuned on-resonance to the wavelength of highest intensity of the incoming pulse. The wavelength spectrum obtained at the drop port shows strong narrowing of the input pulse to about 0.5 nm. One therefore expects a broadened pulse in the time domain. As can be seen in Fig. 7.e, the experiment shows a single largely broadened pulse with a width of 5.2 ps with no fine structure.

In both time spectra, 7.c and e. the intensity has dropped to zero only after about 15 ps. That means that for WDM applications of this MR the bit period should clearly exceed 15 ps in order to avoid inter symbol interference. Bit error rate measurements with the same MR [20] demonstrate error free transmission with a bit-rate of 40 Gbit/sec, return-to-zero (RZ) datasignals corresponding to a bit period of 25 ps. It should be noted that although fulfilling the demands in the time domain of 40 Gbit/s WDM systems, single MRs currently do not perform according to the specifications for, e.g., roll-off, insertion loss or cross-talk.

## 4. Discussion and conclusion

With the theoretical and experimental results in the previous sections a coherent vision on the propagation of short pulses in a microresonator can be given. The main result is that the qualitative features at the output in the time and wavelength domain are strongly dependent on the different characteristic timescales. Most importantly one should mention the integral boundaries of Eq. (1), the time duration Δt. It is for the first time that this important numerical parameter in the conversion of E(t) to E(ν) data is related to an experimentally controllable parameter, namely the characteristic time of the measurement set-up depending on wavelength resolution and the detector speed. Besides Δt, two other time durations determine the qualitative features of the response of a MR to short light pulses: the duration of the pulse, $t_p$ and the roundtrip time of the MR $\tau_r$. There are basically two possibilities: ballistic transport or interference. Ballistic transport and absence of interference is expected if there is *which-way* information for the trajectory from light source to detector. In this case, only replica of the incoming pulse are detected at the output ports of the MR and the pulse can be considered as a quasi particle that propagates ballistically. Condition for this is that $t_p < \tau_r$ and simultaneously $\Delta t < \tau_r$. In all other cases, interference is observed connected to the wave-like properties of light in the pulse. In the time domain, IFD results in pulse broadening whereas in the frequency domain fine structures within the original pulse envelope become visible.

For practical applications in high speed optical communication one often would like to know the bit-rate limit $BR_{limit}$, where a MR can be applied as wavelength selective device. In that case $\tau_r$ should be chosen such that propagation of the data pulses in the ballistic regime can be avoided. Assuming for $t_p$ the duration of a single bit, for example 100 ps for a 10 Gbit/s bit stream, one can write $BR_{limit} = 1/t_{p\ limit}$. The above given condition for IFD $t_p >> \tau_r$ can now be written as $BR_{limit} << 1/\tau_r$. A safe limit for the inequality may be expected when $\tau_r$ is replaced by $F\ \tau_r$. In this way one gets:

$$BR_{limit} = 1/(F\ \tau_r) = \Delta\nu \qquad (9)$$

where $\Delta\nu$ is bandwidth of the MR, i.e. the width at half maximum of the resonance peak. The right-hand identity is based on well-known relations for a MR, see section 2.2. The practical consequences of Eq. (9), which had been derived from the condition for IFD, is in agreement with an intuitive approach stating that the bandwidth of the modulated data stream should fall within the optical bandwidth of the filter.

One can verify the validity of Eq. (9) by performing direct bit-rate error measurements with MRs. In literature several of these measurements with RZ modulation have been reported. Analyzing the presented data one observes in all experiments that the values for error free transmission (BER < $10^{-9}$/s) is always below the limit given by Eq. (9), see Table II. One should note that error free transmission is compatible with slight pulse distortion as evidenced by EYE patterns, see, for example, ref. [20, 22]. On the basis of the 4 experiments referred to in Table II one may conclude that Eq. (9) sets an upper limit and that bitrates beyond this limit should be avoided.

| author | $F$ | $\tau_r$ [ps] | experimental error-free BR [Gbit/s] | $BR_{limit}$ (Eq. 9) |
|---|---|---|---|---|
| Suzuki et al. [21] | 12 | 0.5 | 50 | 170 |
| Geuzebroek et al. [22] | 10 | 1.9 | 40 | 53 |
| Geuzebroek et al. [20] | 23 | 1.9 | 10 | 22 |
| Melloni et al. [23] | 7.5 | 10 | 10 | 13 |

*Table II: Comparison of experimental error free transmission of optical data through MRs with bit-rate limits obtained with the aid of Eq. (9).*

**Acknowledgments**
The authors would like to thank H-J Gersen of UT-MESA[+] and Niehusmann of the RWTH Aachen for collaboration in part of the measurements and Lucy T.H. Hilderink, Henry Kelderman and Gabriel Sengo for their contribution in the realization of the devices. We would like to acknowledge discussions with Hugo Hoekstra, Dion Klunder, Niek van Hulst and Kobus Kuipers. This work has financially been supported by EC project NAIS, the NoE ePIXnet and the Dutch BSIK Freeband Broadband Photonics project.

**Figures**

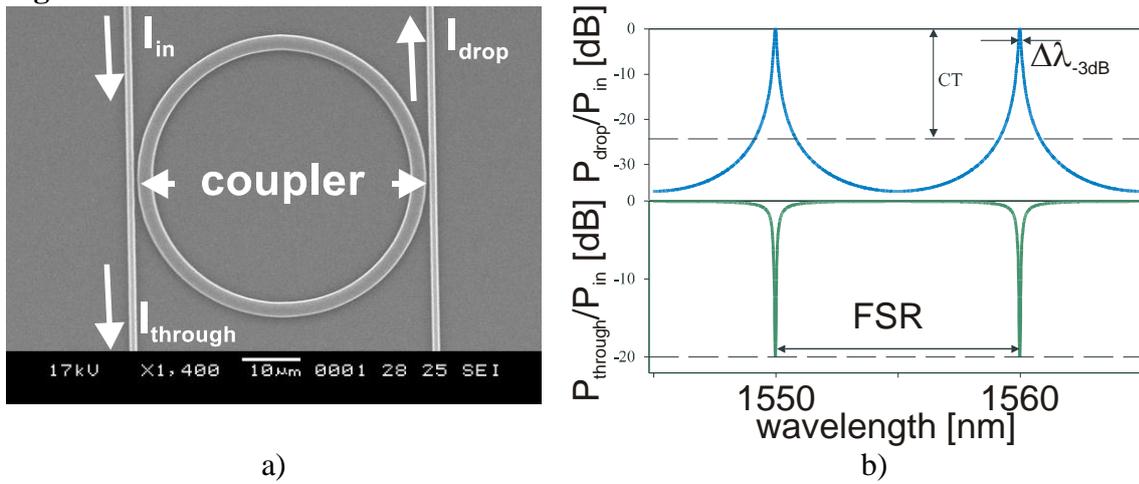

a)                                        b)

*Figure 1. MR with two adjacent waveguides serving as in- and output port; (a) SEM picture of device (⌀ = 50 μm);( b) relative power in the drop and through port of a loss-less MR with F ~ 100 as a function of wavelength.*

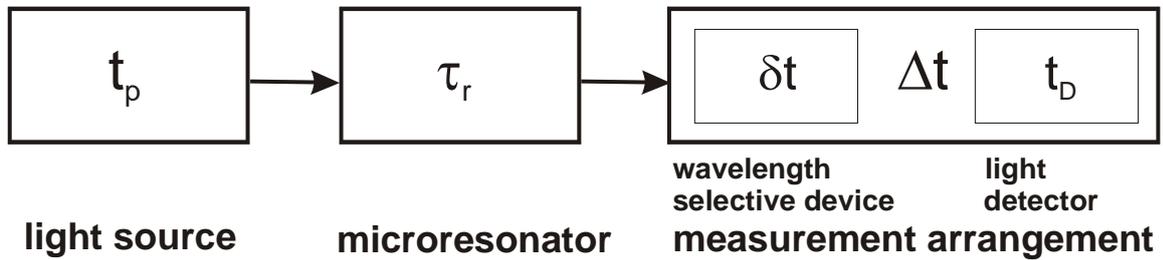

*Fig. 2  Schematic set-up for the measurement of propagation of short pulses through an optical MR*

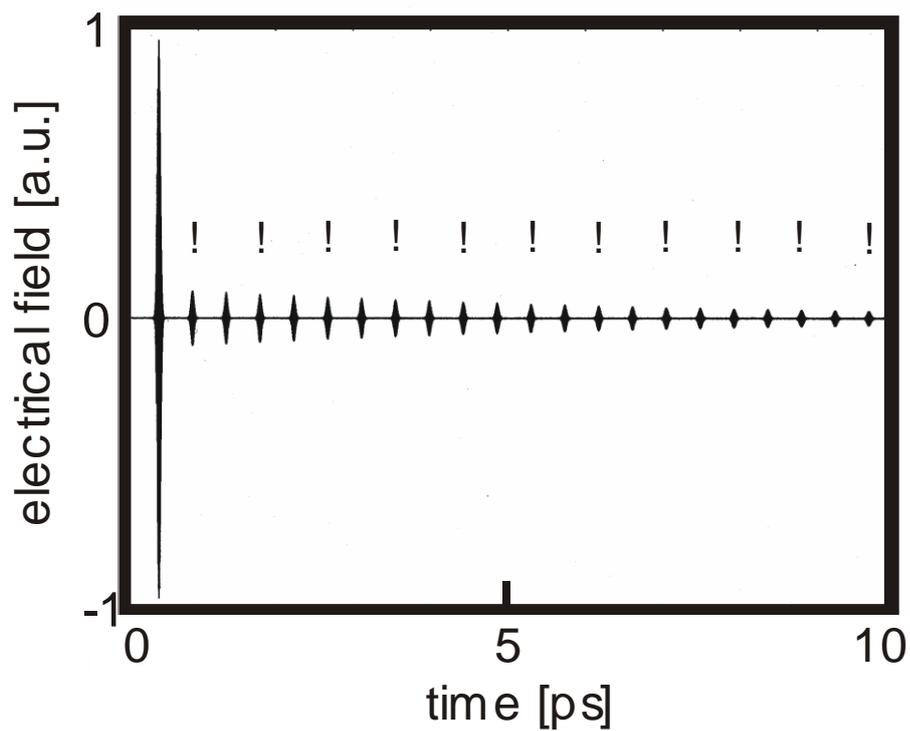

*Fig. 3 Calculated field response to a 20 fs single pulse at the through and drop ports. The bars indicate the pulse train obtained at the drop port.*

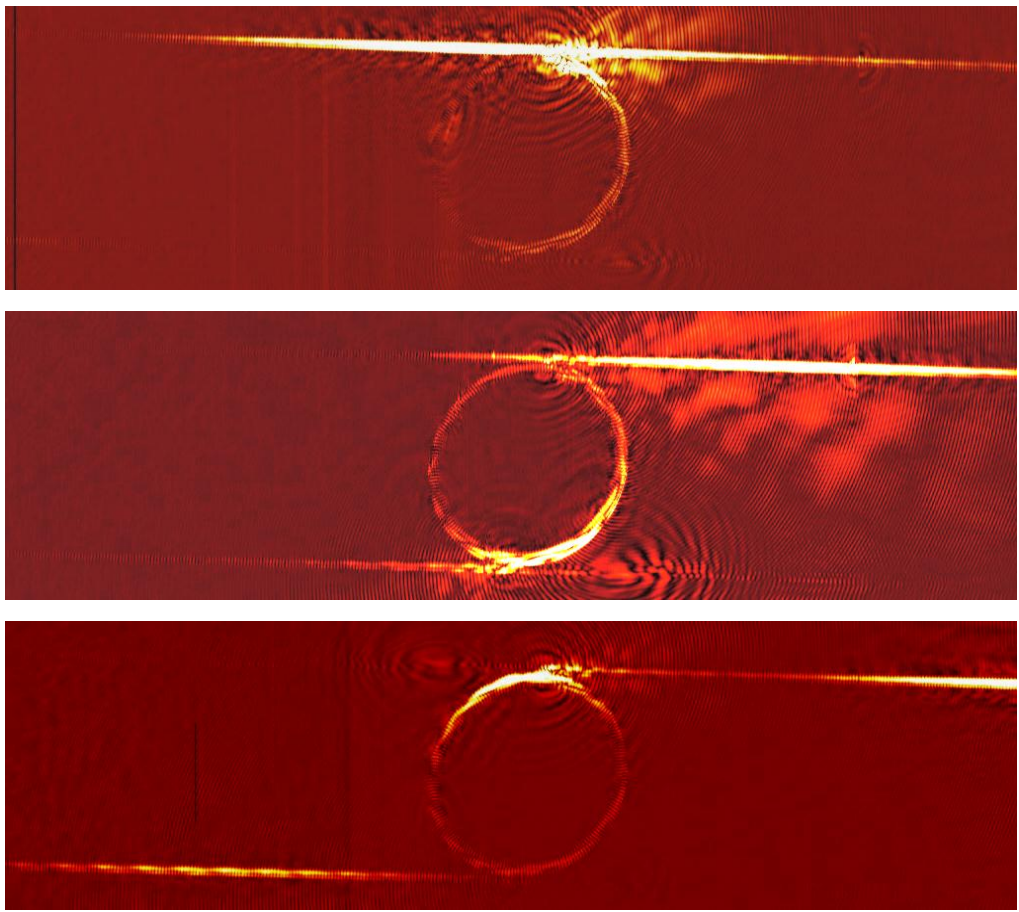

*Fig. 4. Snapshot of the electromagnetic field of a fs lightpulse in a microresonator (⌀ = 25 μm) with port waveguides; a) pulse just partly entering the ring, b) 400 fs later, c) 1000 fs later.*

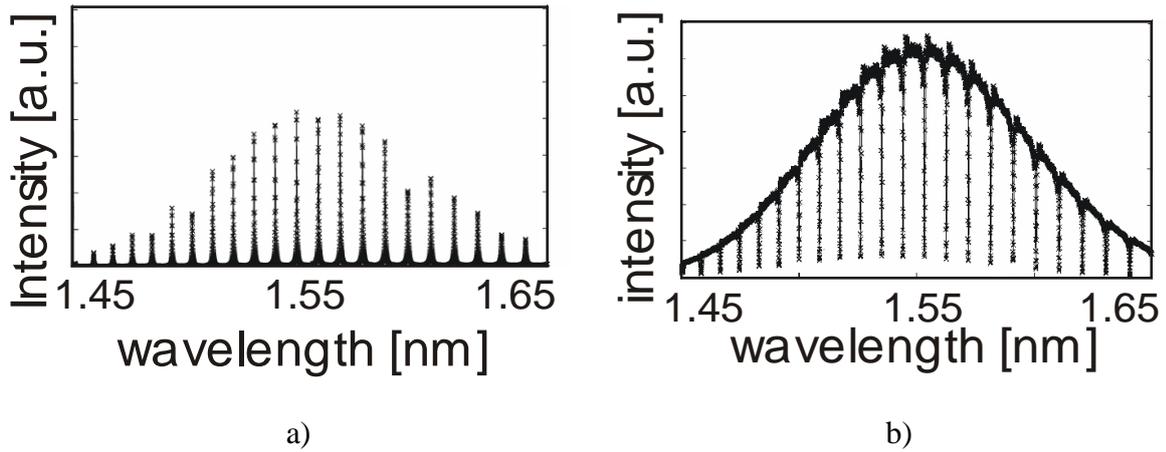

*Fig. 5: Spectral response of the microresonator obtained by Fourieranalysis of the output fields (see Fig. 3): a) through port, b) drop port.*

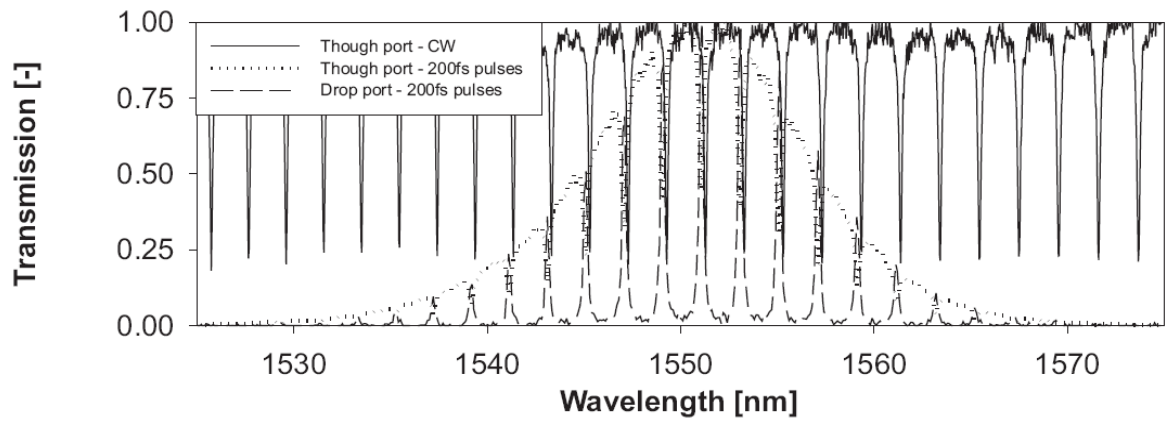

*Fig. 6: Experimental wavelength spectrum after propagation of CW and 200 fs pulses through MR2.*

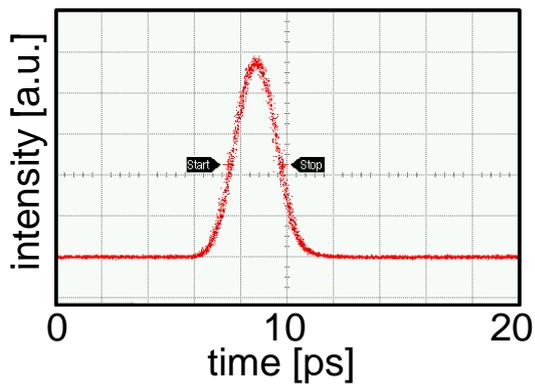

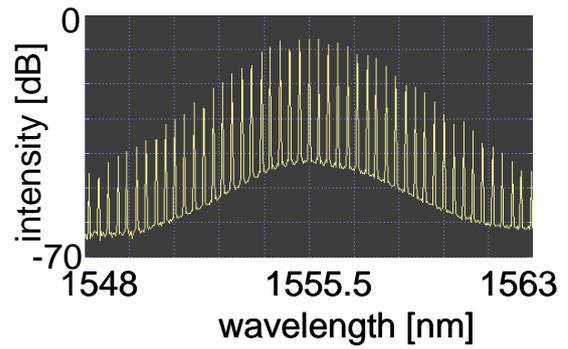

(a) (b)

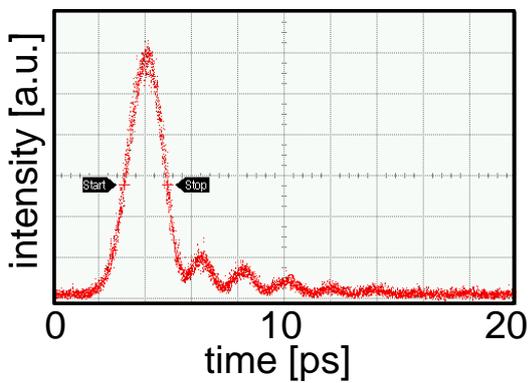

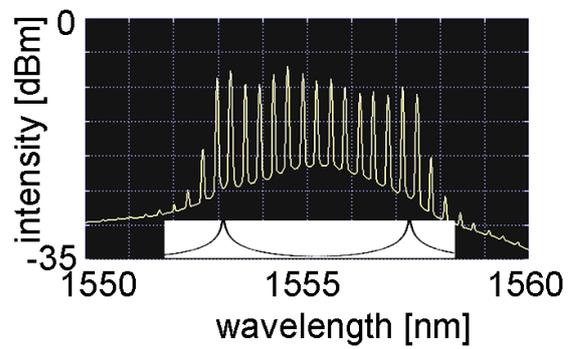

(c) (d)

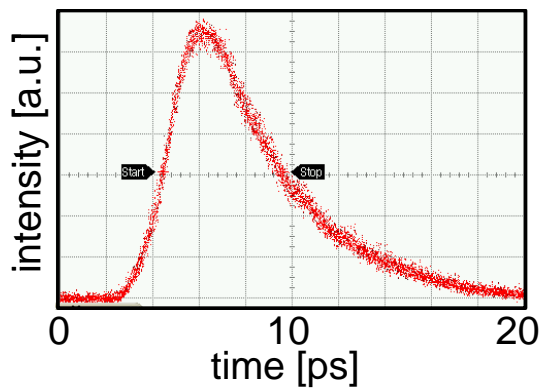

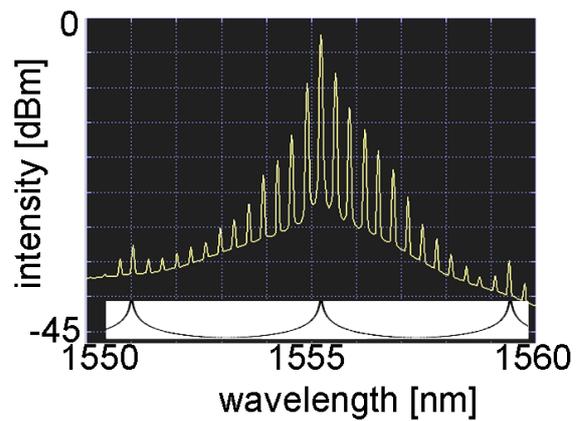

(e) (f)

*Fig. 7 Response at the drop port of a MR to a laser pulse of duration ~2 ps with a repetition rate of 40 GHz; (a) time and (b) wavelength spectrum of the incoming pulse; time (c) and wavelength (d) response with MR tuned off-resonance to the center wavelength of the laser; time (e) and wavelength (f) response with MR tuned on-resonance. In the insert of (d) and (f) the transmission function of the MR with the position of the resonance fringes is shown schematically.*